\documentclass[twocolumn]{aastex63}
\usepackage{newtxtext,newtxmath}
\usepackage{graphicx}
\usepackage{mathrsfs}

\newcommand{\kms}{km\,s$^{-1}$}
\newcommand{\cmsq}{cm$^{-2}$}

\newcommand{\hi}{\mbox{H\,{\sc i}}}
\newcommand{\hii}{\mbox{H\,{\sc ii}}}

\shorttitle{21cm and optical line mapping of a z=0.026 DLA}
\shortauthors{Boettcher et al.}

\begin{document}

\title{Discovery of a damped Ly$\alpha$ absorber originating in a spectacular interacting dwarf galaxy pair at $z = 0.026$\footnote{Based on data gathered with the 6.5m Magellan Telescopes located at Las Campanas Observatory and the NASA/ESA Hubble Space Telescope operated by the Space Telescope Science Institute and the Association of Universities for Research in Astronomy, Inc., under NASA contract NAS 5-26555.}}

\correspondingauthor{Erin Boettcher}
\email{eboettch@umd.edu}

\author{Erin Boettcher}
\affiliation{Department of Astronomy, University of Maryland, College Park, MD 20742, USA}
\affiliation{Code 662, NASA Goddard Space Flight Center, Greenbelt, MD 20771, USA}
\affiliation{Department of Astronomy \&
  Astrophysics, The University of Chicago, 5640 S. Ellis Ave.,
  Chicago, IL 60637, USA}

\author{Neeraj Gupta}
\affiliation{Inter-University Centre for Astronomy and Astrophysics, Post Bag 4, Ganeshkhind, Pune 411 007, India}

\author{Hsiao-Wen Chen}
\affiliation{Department of Astronomy \&
  Astrophysics, The University of Chicago, 5640 S. Ellis Ave.,
  Chicago, IL 60637, USA}

\author{Mandy C. Chen}
\affiliation{Department of Astronomy \&
  Astrophysics, The University of Chicago, 5640 S. Ellis Ave.,
  Chicago, IL 60637, USA}

\author{Gyula I. G. J\'ozsa}
\affiliation{South African Radio Astronomy Observatory, 2 Fir Street, Black River Park, Observatory, Cape Town, 7925, South Africa}
\affiliation{Department of Physics and Electronics, Rhodes University, P.O. Box 94, Makhanda, 6140, South Africa}
\affiliation{Argelander-Institut f\"{u}r Astronomie, Auf dem H\"{u}gel 71, D-53121 Bonn, Germany}

\author{Gwen C. Rudie}
\affiliation{The Observatories of the Carnegie Institution for Science, 813 Santa Barbara Street, Pasadena, CA 91101, USA}

\author{Sebastiano Cantalupo}
\affiliation{Department of Physics, University of Milan Bicocca, Piazza della Scienza 3, 20126 Milano, Italy}

\author{Sean D. Johnson}
\affiliation{Department of Astronomy,
  University of Michigan, Ann Arbor, MI 48109, USA}

\author{S. A. Balashev}
\affiliation{Ioffe Institute, Politekhnicheskaya 26, 194021 Saint Petersburg, Russia}
\affiliation{HSE University, Saint Petersburg, Russia}

\author{Fran\c{c}oise Combes}
\affiliation{Observatoire de Paris, LERMA, Coll\`{e}ge de France, CNRS, PSL University, Sorbonne University, 75014 Paris, France}

\author{Kathy L. Cooksey}
\affiliation{Department of Physics and Astronomy, University of Hawai’i at Hilo, Hilo, HI 96720, USA}

\author{Claude-Andr\'{e} Faucher-Gigu\`{e}re}
\affiliation{Department of Physics \& Astronomy and Center for Interdisciplinary Exploration and Research in Astrophysics (CIERA), Northwestern University, 1800 Sherman Ave, Evanston, IL 60201, USA}

\author{Jens-Kristian Krogager}
\affiliation{Department of Astronomy, University of Geneva, Chemin Pegasi 51, 1290 Versoix, Switzerland}

\author{Sebastian Lopez}
\affiliation{Departamento de Astronom\'{i}a, Universidad de Chile, Casilla 36-D, Santiago, Chile}

\author{Emmanuel Momjian}
\affiliation{National Radio Astronomy Observatory, P. O. Box O, Socorro, NM 87801, USA}

\author{Pasquier Noterdaeme}
\affiliation{Institut d’Astrophysique de Paris, CNRS-SU, UMR7095, 98bis Bd Arago, 75014 Paris, France}
\affiliation{Franco-Chilean Laboratory for Astronomy, IRL 3386, CNRS and Departamento de Astronom\'ia, Universidad de Chile, Casilla 36-D, Santiago, Chile}

\author{Patrick Petitjean}
\affiliation{Institut d’Astrophysique de Paris, CNRS-SU, UMR7095, 98bis Bd Arago, 75014 Paris, France}

\author{Marc Rafelski}
\affiliation{Space Telescope Science Institute, Baltimore, MD 21218, USA}
\affiliation{Department of Physics \& Astronomy, Johns Hopkins University, Baltimore, MD 21218, USA}

\author{Raghunathan Srianand}
\affiliation{Inter-University Centre for Astronomy and Astrophysics, Post Bag 4, Ganeshkhind, Pune 411 007, India}

\author{Gregory L. Walth}
\affiliation{The Observatories of the Carnegie Institution for Science, 813 Santa Barbara Street, Pasadena, CA 91101, USA}

\author{Fakhri S. Zahedy}
\affiliation{The Observatories of the Carnegie Institution for Science, 813 Santa Barbara Street, Pasadena, CA 91101, USA}



\begin{abstract}

We present the discovery of neutral gas detected in both damped
Ly$\alpha$ absorption (DLA) and \hi\ 21-cm emission outside of the
stellar body of a galaxy, the first such detection in the literature.
A joint analysis between the Cosmic Ultraviolet Baryon Survey and the
MeerKAT Absorption Line Survey reveals an \hi\ bridge connecting two
interacting dwarf galaxies (log$(M_{\text{star}}/\text{M}_{\odot}) =
8.5 \pm 0.2$) that host a $z = 0.026$ DLA with log[$N$(\hi)/cm$^{-2}$]
$ = 20.60 \pm 0.05$ toward the QSO J2339$-$5523 ($z_{\text{QSO}} =
1.35$). At impact parameters of $d = 6$ and $33$ kpc, the dwarf
galaxies have no companions more luminous than $\approx 0.05L_{*}$
within at least $\Delta\,\varv = \pm 300$ km s$^{-1}$ and $d \approx
350$ kpc. \hi\ 21-cm emission is spatially coincident with the DLA at
the 2--3$\sigma$ level per spectral channel over several adjacent
beams. However, \hi\ 21-cm absorption is not detected against the
radio-bright QSO; if the background UV and radio sources are spatially
aligned, the gas is either warm or clumpy (with spin temperature to
covering factor ratio $T_{s}/f_{c} > 1880$ K). VLT-MUSE observations
demonstrate that the $\alpha$-element abundance of the ionized ISM is
consistent with the DLA ($\approx 10$\% solar), suggesting that the
neutral gas envelope is perturbed ISM gas. This study showcases the
impact of dwarf-dwarf interactions on the physical and chemical state
of neutral gas outside of star-forming regions. In the SKA era, joint
UV and \hi\ 21-cm analyses will be critical for connecting the cosmic
neutral gas content to galaxy environments.

\end{abstract}




\section{Introduction}
\label{sec:intro}

Damped Lyman-$\alpha$ absorbers (DLAs; log[$N$(\hi)/cm$^{-2}$] $\ge
20.3$) have sufficiently large \hi\ column densities to be
self-shielded against ionizing radiation and trace a dominant fraction
of neutral gas in the Universe, making them critical probes of the
galactic and circumgalactic environments related to star formation
\citep[for reviews, see][]{2005ARA&A..43..861W,
  2017ASSL..434..291C,2017MNRAS.469.2959K}. Low-redshift DLAs ($z<1$)
detected through ultraviolet (UV) absorption-line spectroscopy using
the unique capabilities of the \textit{Hubble Space Telescope}
(\textit{HST}) are particularly important to our understanding of DLA
host galaxies, which can be most easily detected and characterized in
the local Universe (e.g., \citealt{2012ApJ...744...93B,
  2016ApJ...818..113N}; though note discussion of possible redshift
evolution in the DLA host galaxy population, e.g.,
\citealt{2019ApJ...870L..19N, 2021ApJ...921...68K}). Previous studies
suggest that low-redshift DLA host galaxies resemble the general field
population and are thus diverse in their colors, luminosities, and
environments, including both isolated galaxies and group members
\citep[e.g.,][]{2003ApJ...597..706C, 2005ApJ...620..703C,
  2011MNRAS.416.1215R, 2016MNRAS.457..903P}.

Ambiguities in the origin of DLA gas --- whether it is from a gaseous
disk, halo, intragroup gas, or infalling/outflowing material --- can
be understood by observing \hi\ in emission through the 21-cm
transition, which is sensitive to $N($\hi$) = \text{a few} \times
10^{18}$ cm$^{-2}$. If the background quasar is also bright at radio
wavelengths, the \hi\ 21-cm absorption can be used to constrain the
cold neutral medium (CNM) fraction of the DLA and the clumpiness of
any cold gas \citep[e.g.,][]{Srianand12dla, Kanekar14}. However, DLAs
in front of radio-loud quasars and at sufficiently low redshift
($z<0.1$) to be detectable in \hi\ 21-cm emission with current radio
telescopes remain extremely rare.

\citet{2018MNRAS.473L..54K} report \hi\ 21-cm emission-line
observations of four DLAs at $z < 0.1$, but the spatial resolution of
these single-dish data is inadequate to unambiguously associate the
21-cm emission to individual galaxies or to dissect its origin. To the
best of our knowledge, only two cases of bonafide DLAs with spatially
resolved \hi\ 21-cm observations have been reported in the
literature. In the first case, the quasar sightline passes through the
optical extent of a low surface brightness galaxy at $z = 0.009$
\citep{2001A&A...372..820B, 2002A&A...388..383C}. In the second case,
\citet{2019ApJ...871..239B} mapped the neutral gas properties of a
galaxy group that hosts a DLA at $z=0.029$; no \hi\ 21-cm emission was
detected to be spatially coincident with the DLA. The quasar in both
cases is not bright at radio wavelengths. \hi\ 21-cm emission-line
detections of several galaxies associated with 21-cm absorbers have
also been reported in the literature \citep[e.g.,][]{Carilli92,
  Dutta16, Gupta18j1243}. These confirm the complexity of the
galaxy/absorber relationship, but UV-selected DLAs with spatially
resolved \hi\ 21-cm emission-line mapping of their galactic
environments remain scarce.

Here, we report the rare case of a bonafide DLA ($z = 0.026$) that
coincides with \hi\ 21-cm emission outside of the optical extent of a
galaxy. The background QSO J2339$-$5523 ($z_{\text{QSO}} = 1.35$) is
also bright at radio wavelengths (185\,mJy at 1.4\,GHz), and thus we
can constrain the thermal state and clumpiness of the gas through
\hi\ 21-cm absorption. A deep galaxy redshift survey demonstrates that
the DLA is associated with an interacting dwarf galaxy pair that is
otherwise isolated from massive companions. This system is a valuable
case study of the impact of dwarf-dwarf interactions on their neutral
gas reservoirs independent of the influence of a nearby massive galaxy
and its hot halo.

Throughout the paper, we adopt the solar abundance pattern of
\cite{Asplund2009}. We assume a $\Lambda$ cosmology with
$\Omega_{\text{M}} = 0.3$, $\Omega_{\Lambda} = 0.7$, and $H_{0} = 70$
km\,s$^{-1}$ Mpc$^{-1}$.

\section{Data and Analysis}
\label{sec:data}

This study is based on data from the Cosmic Ultraviolet Baryon Survey
(CUBS; \citealt{Chen2020}) and the MeerKAT Absorption Line Survey
(MALS; \citealt{Gupta2016}). Below we describe the data reduction and
analysis of UV absorption-line spectroscopy of J2339-5523 in
\S\ref{sec:COS_analysis}, \hi\ 21-cm observations from MeerKAT in
\S\ref{sec:MeerKAT}, and optical integral field unit (IFU)
observations from the Multi Unit Spectroscopic Explorer (MUSE) on the
Very Large Telescope in \S\ref{sec:MUSE}.

\subsection{UV Absorption-Line Analysis}
\label{sec:COS_analysis}

We refer the reader to \citet{Chen2020} for a description of the
absorption-line observations of J2339-5523 from the Cosmic Origins
Spectrograph (COS) on \textit{HST} ($\lambda = 1100\text{--}1800$ \AA,
spectral resolution FWHM $\approx 20$ km s$^{-1}$) and the Magellan
Inamori Kyocera Echelle (MIKE) spectrograph on Magellan-Clay
($\lambda$ = 3200\text{--}9200 \AA, FWHM $\approx 8$ km s$^{-1}$). We
use Voigt profile fitting to determine the velocity ($\varv$), Doppler
parameter ($b$), and column density ($N$) of the atomic and ionic
transitions associated with the DLA using a grid search as described
in \S3 of \citet{2021ApJ...913...18B}. The COS wavelength calibration
is sufficiently uncertain that we allow small variations within
$\delta\varv = \pm 5$ km s$^{-1}$ for low and intermediate ions in COS
with respect to \ion{Ca}{2}$\lambda\lambda$3934, 3969 in MIKE. We
construct a marginalized probability density function for each
parameter by calculating a likelihood of the form $\mathscr{L} \propto
{\rm e}^{-\chi^{2}/2}$ at every point in parameter space, normalizing
the total likelihood to unity, and integrating over the marginalized
parameters. We adopt as the best-fit model the parameters
corresponding to a value of 50\% from the cumulative distribution
function. The reported uncertainties correspond to the 68\% confidence
interval, and the upper limits are reported as the 95\% one-sided
confidence interval. We additionally perform a curve of growth (COG)
analysis for the low and intermediate ions to corroborate the
characteristic Doppler parameter determined by the Voigt profile
fitting.

\subsection{MeerKAT \hi\ 21-cm Analysis}
\label{sec:MeerKAT}

We observed the field centered at J2339-5523 using the MeerKAT-64
array on 2020 June 14 and 22. Of the 64 antennas, 59 and 60 antennas
participated in the first and second observing run, respectively. We
used the 32K mode of the SKA Reconfigurable Application Board (SKARAB)
correlator to split the total bandwidth of 856 MHz centered at
1283.9869 MHz into 32,768 frequency channels. The resultant frequency
resolution is 26.123\,kHz, or 5.7\,\kms, at the redshifted \hi\ 21-cm
line frequency of the DLA. PKS\,1939-638 and PKS\,0408-658 were
observed for flux density, delay, and bandpass calibrations, and the
compact radio source J2329-4730 was observed for complex gain
calibration. The total on-source time on J2339-5523 was 112\,mins.

The MeerKAT data were processed using the Automated Radio Telescope
Imaging Pipeline (ARTIP); we refer the reader to \citet{Gupta21} for
details.  Here, we focus on the Stokes-$I$ radio continuum and
spectral line properties near the \hi\ 21-cm line frequency
corresponding to $z_{\text{DLA}}$ ($\sim$1384.4\,MHz). The spatial
resolution represented by the synthesized beam of the continuum image
and image cube obtained using {\tt robust = 0} weighting is
$8.4^{\prime\prime}\times6.6^{\prime\prime}$ (position angle =
13$^\circ$.8), with a spatial pixel size of $2.0^{\prime\prime}$. The
cube has been deconvolved, or cleaned, using the CASA task {\tt
  tclean} down to five times the single channel rms of
0.45\,mJy\,beam$^{-1}$. The radio emission associated with J2339-5523
is compact with a deconvolved source size of $<0.6^{\prime\prime}$.

We used the \hi\ Source Finding Application \citep[SOFIA
  v2.0;][]{Serra15, Westmeier21} for the \hi\ 21-cm analysis of the
spectral line cube.  We set SOFIA to subtract residual continuum
subtraction errors from the image cube by fitting a polynomial of
order 1. We used the smooth+clip (S+C) algorithm with a combination of
spatial kernels of 0, 3, 6, and 9 pixels, and spectral kernels of 0,
3, 7, and 15 channels, to detect voxels containing \hi\ 21-cm emission
in the image cube.  We set the threshold of the S+C finder to
3.5$\sigma$ and used the reliability filter to reject unreliable
detections.  The output three dimensional mask was then used to
further clean the image cube down to the single channel rms.  This
deep cleaned cube was then passed through SOFIA with the
above-mentioned setup to generate the final \hi\ moment maps and the
integrated spectrum. For the moment 1 and 2 maps, we used only those pixels
where the signal was detected at 1.5 times the local rms. In the
moment 0 map, the mean value of pixels with $S/N$ in the range
$2.5$--$3.5\sigma$ corresponds to an \hi\ column density of $N$(\hi) =
$3.2 \times 10^{20}$ cm$^{-2}$. This represents the average $3\sigma$
column density sensitivity across the map (see
\citealt{2018ApJ...865...26D} for details of this sensitivity
estimate).

\subsection{VLT-MUSE Analysis}
\label{sec:MUSE}

We observed the field of J2339-5523 on 2020 Nov 22, Dec 05, and Dec 09
using VLT-MUSE. The total exposure time on source was 9390\,s. We
reduced the data using the standard ESO MUSE pipeline
\citep{2014ASPC..485..451W} and the custom software CubExtractor from
S. Cantalupo \citep{2019MNRAS.483.5188C}. We subtracted the quasar
light using a high-resolution spectral differential imaging method
\citep[e.g.,][]{2019NatAs...3..749H}. The wavelength coverage is
$\lambda = 4700 - 9350$ \AA, and the spectral resolution at the blue
(red) end is $R \approx 1610$ ($R \approx 3570$). This corresponds to
a FWHM of $\approx 185$ km s$^{-1}$ ($85$ km s$^{-1}$), with
approximately two wavelength bins sampling the spectral resolution
element. Multiple pointings mapped a region of $1.3' \times 1.4'$
around the QSO. We spatially smoothed the MUSE data cube with a boxcar
kernel of $3 \times 3$ pixels, where the kernel size was chosen to
approximate the seeing disk at the time of the observations ($\approx
0.6''$, or $0.3$ kpc at the redshift of the DLA).

We construct continuum-subtracted narrow-band images at the wavelength
of all emission lines at the redshift of the DLA. We subtract the
continuum by fitting a linear function locally around the masked
lines. In parts of both dwarf galaxies, Balmer absorption from the
stellar continuum modestly impacts the observed H$\alpha$ and H$\beta$
emission. To correct for this, we use the stellar population synthesis
code
\texttt{bagpipes}\footnote{https://bagpipes.readthedocs.io/en/latest/}
\citep{Carnall2018} to fit a tau model to the stellar continuum of a
bright, star-forming region in each galaxy assuming a spatially
uniform star-formation history; the equivalent widths of the Balmer
lines in the best-fit models are in the range $2.5$--$3.0$ \AA. We
then scale the models to the continuum level of any spaxel for which
the continuum light is detected at $\ge 3\sigma$ at the wavelength of
H$\alpha$ or H$\beta$. To allow for a possible velocity offset between
the stars and the gas, we adopt a broad window of $\pm 500$ km
s$^{-1}$ ($\pm 180$ km s$^{-1}$) when constructing the narrow-band
images of the Balmer lines for spaxels with (without) this correction
applied. We derive an extinction correction using the extinction law
of \citet{1989ApJ...345..245C} from the reddening, $E(B - V)$,
calculated from the Balmer decrement following \citet{Calzetti1994}
for every spaxel for which H$\beta$ and H$\alpha$ are detected at the
$\ge 3\sigma$ level.

Finally, we fit the six strongest emission lines (H$\alpha$, H$\beta$,
[OIII]$\lambda\lambda$4960, 5008, and [SII]$\lambda\lambda$6718, 6732)
with a single Gaussian profile using the \texttt{IDL} software
\texttt{KUBEVIZ}\footnote{http://www.mpe.mpg.de/~dwilman/kubeviz/}. This
yields a velocity centroid and velocity dispersion, corrected for the
instrumental resolution, for the ionized gas in each spaxel with
statistically significant emission.

\section{Results}
\label{sec:results}

We present the properties of the DLA in \S\ref{sec:DLA_prop}. We then
discuss the associated dwarf galaxy pair in \S\ref{sec:dwarfs}, before
characterizing the neutral and warm ionized gas properties of the
dwarfs and their environs in the following two Sections.

\subsection{A dusty, low-metallicity DLA}
\label{sec:DLA_prop}

\begin{figure*}
\centering
\includegraphics[scale = 0.85]{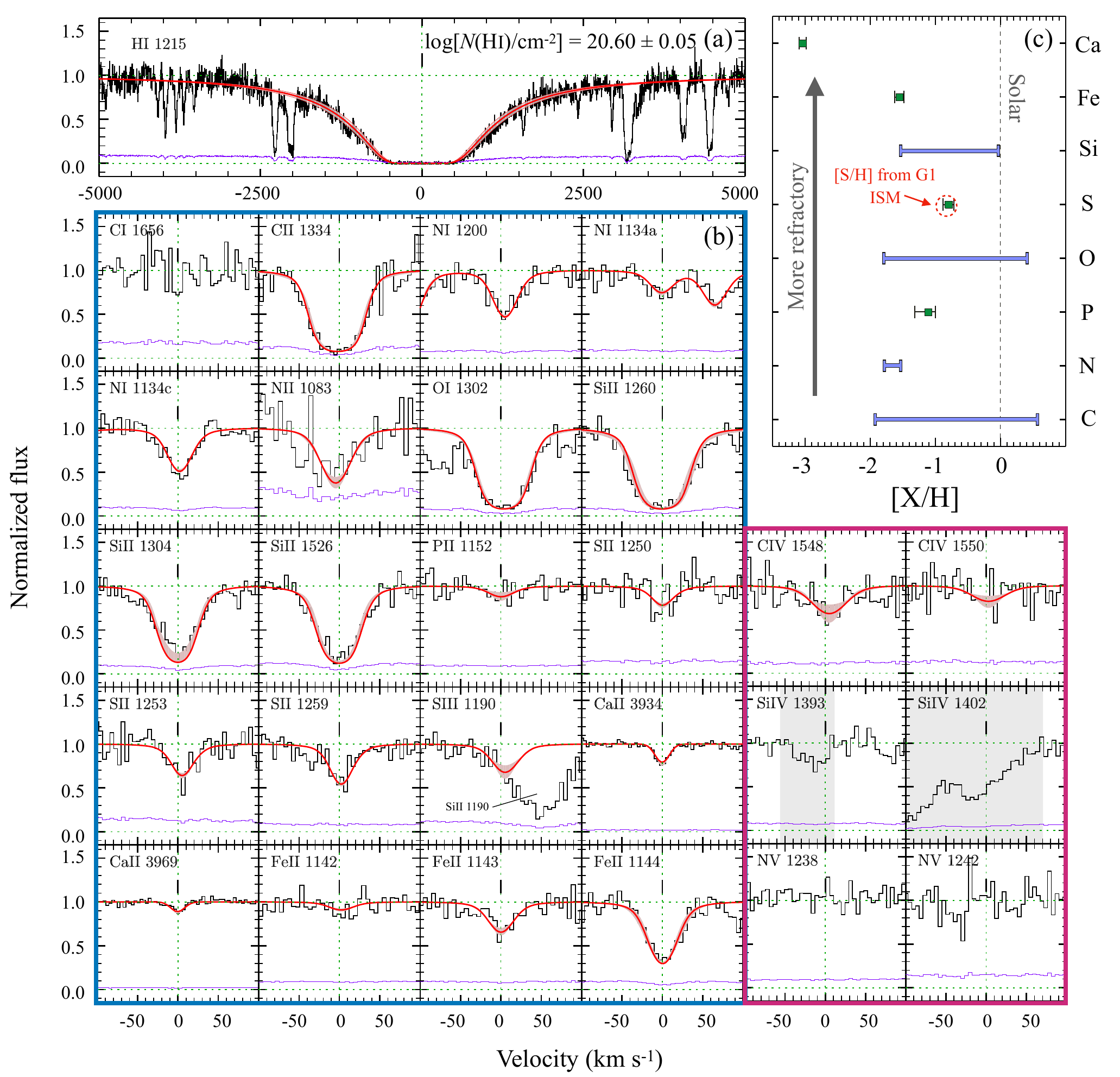}
\caption{\textit{HST}-COS and Magellan-MIKE absorption-line
  spectroscopy of the $z = 0.026$ DLA towards J2339-5523. \textbf{(a),
    (b)} The data, error, and best-fit models are shown in black,
  purple, and red, respectively, for Ly$\alpha$ and a representative
  sample of low (left, blue box) and intermediate (right, pink box)
  ions. The gray shaded regions around the best-fit models represent
  the 1$\sigma$ uncertainty. For ions without independent constraints
  on $b$, the red line is the model with $b = 15$ km s$^{-1}$ and the
  shaded region shows the range of models with $10$ km s$^{-1}$ $\leq
  b \leq 20$ km s$^{-1}$. The COS data are binned by two spectral
  pixels, or $\Delta\varv \approx 5$ km s$^{-1}$, and the velocity
  zeropoint is the DLA velocity. \textbf{(c)} The chemical abundance,
  [X/H], derived for a range of elements ordered from most volatile
  (bottom) to most refractory (top). Elements with well-constrained
  abundances are indicated in green. The metallicity of the gas is
  determined to be $Z \approx 0.1Z_{\odot}$ by [S/H]. The decline
  towards lower [X/H] for the most refractory elements suggests the
  presence of dust. The red dashed circle shows the $1\sigma$ range of
  [S/H] measured for the brightest star-forming region in the more
  proximate dwarf, demonstrating strong consistency in the
  $\alpha$-element abundances of the damped absorber and the ISM of
  the host galaxy.}
\label{fig:COS}
\end{figure*}

\begin{deluxetable*}{lccc}
  \tablecaption{$z = 0.026$ DLA Properties}
  \tablehead{
    \multicolumn{4}{c}{\textbf{Column densities for limits and ions fit with $b$ as free parameter}\tablenotemark{a}} \\    
    \colhead{Ion} &
    \colhead{$b$ (km\,s$^{-1}$)} & 
    \colhead{log[$N$/cm$^{-2}$]} &
    \colhead{Comments}
  }
  \startdata
  \noalign{\smallskip}
  \ion{H}{1} & $--$& $20.60 \pm 0.05$ & $b$ unconstrained due to Ly$\alpha$ damping wings; $N$ independent of $b$ \\        
  \ion{C}{1} & $--$ & $< 13.4$ & $b$ fixed to $10$ \kms\ based on \ion{Ca}{2} \\
  \ion{C}{2*} & $--$ & $< 13.3$ & $b$ fixed to $10$ \kms\ based on \ion{Ca}{2} \\
  \ion{C}{4} & $23\substack{+12 \\ - 8}$ & $13.45 \pm 0.08$ \\
  \ion{N}{1} & $12 \pm 2$ & $14.46\substack{+0.05 \\ -0.04}$ \\
  \ion{N}{5} & $--$ & $< 12.9$ & $b$ fixed to $25$ \kms\ based on \ion{C}{4}; comparable $N$ found for $b \lesssim 100$ \kms \\
  \ion{Si}{4} & $--$ & $< 13.0$ & $b$ fixed to $25$ \kms\ based on \ion{C}{4}; both 1393 and 1402 \AA\ transitions contaminated  \\
  \ion{P}{2} & $--$ & $12.9\substack{+0.1 \\ - 0.2}$ & Probability density function is flat in $b$ for $10$ \kms\ $\leq b \leq 20$ \kms\ \\
  \ion{S}{2} & $10\substack{+2 \\ -1}$ & $14.77\substack{+0.07 \\ - 0.05}$ \\
  \ion{S}{3} & $--$ & $14.4 \pm 0.1$ & \ion{S}{3}$\lambda$1190 blended with \ion{Si}{2}$\lambda$1190; \ion{S}{3} $b$ value not well constrained \\    
  \ion{Ca}{2} & $10.3\substack{+0.8 \\ -0.7}$ & $11.9 \pm 0.02$ \\
  \ion{Fe}{2} & $14\substack{+4 \\ -3}$ & $14.55 \pm 0.05$ \\    
  \noalign{\smallskip}
  \hline\hline
  \multicolumn{4}{c}{\textbf{Column densities for $b = 10$ \kms, $b = 20$ \kms}\tablenotemark{b}} \\
   & log[$N$/cm$^{-2}$] & log[$N$/cm$^{-2}$] & \\
   & ($b = 10$ \kms) & ($b = 20$ \kms) & \\
  \hline
  \ion{C}{2} & 17.6 & 15.1 & \ion{C}{2}$\lambda$1334 is saturated \\
  \ion{N}{2} & 14.7 & 14.2 & Only low $S/N$ \ion{N}{2}$\lambda$1083 transition available \\
  \ion{O}{1} & 17.7 & 15.5 & \ion{O}{1}$\lambda$1302 is saturated \\
  \ion{Al}{2} & 14.9 & 13.3 & \ion{Al}{2}$\lambda$1670 is saturated \\
  \ion{Si}{2} & 15.9 & 14.5 & Five saturated lines available \\
  \ion{Si}{3} & 15.6 & 13.8 & \ion{Si}{3}$\lambda$1206 is saturated
  \label{tab:metals}
  \enddata
  \tablenotetext{a}{We impose the prior that $10$ \kms\ $\leq b \leq 20$ \kms\ in the Voigt profile fitting of neutral, singly, and doubly ionized species based on the COG analysis.}
  \tablenotetext{b}{We report the best-fit column density corresponding to $b = 10$ \kms\ and $b = 20$ \kms\ for ions with only saturated and/or low $S/N$ transitions available; the $b$ values are chosen based on the COG analysis.}
\end{deluxetable*}

In Table \ref{tab:metals}, we list the best-fit column densities and
Doppler parameters for all atomic and ionic species associated with
the DLA, and we show a representative sample of the absorption lines
and their best-fit models in Fig.~\ref{fig:COS}. The \hi\ column
density determined from the Ly$\alpha$ line is log[$N$(\hi)/cm$^{-2}$]
= $20.60 \pm 0.05$. The uncertainty is dominated by the continuum
fitting, as determined by repeated measurements of $N$(\hi) from
independent, manual fits to the local continuum prior to the
Ly$\alpha$ fitting.

The \ion{Ca}{2}$\lambda\lambda$3934, 3969 lines are the only
associated absorption features detected in the MIKE data and indicate
a single velocity component at $z = 0.02604$, which we adopt as the
velocity zeropoint in the following analysis. The \ion{Ca}{2} lines
have a Doppler parameter of $b \approx 10$ km s$^{-1}$, which we
corroborate using a COG analysis for low and intermediate ions in COS
that span a wide range of oscillator strengths. The best constraints
come from six \ion{Fe}{2} and six \ion{N}{1} transitions; these
measurements suggest that $10\ \text{km s$^{-1}$} \le b \le 20$ km
s$^{-1}$, and we impose this prior in the Voigt profile fitting of
neutral, singly, and doubly ionized species. Only saturated or low
$S/N$ lines are available for \ion{O}{1}, \ion{C}{2}, \ion{N}{2},
\ion{Al}{2}, \ion{Si}{2}, and \ion{Si}{3}, and we report a range of
column densities corresponding to $10\ \text{km s$^{-1}$} \le b \le
20$ km s$^{-1}$ for these species. There is some indication of
moderate ionization state gas associated with the DLA from the
detection of \ion{C}{4} (the \ion{Si}{4} transitions are contaminated
by interlopers). The \ion{C}{4} Doppler parameter exceeds that of the
low ions by a factor of two, hinting that it may arise in a boundary
layer between the neutral absorber and a hot ambient medium
\citep[e.g.,][]{2005ApJ...630..332F}.

In panel (c) of Fig.~\ref{fig:COS}, we show the chemical abundances,
[X/H], for a range of elements ordered from most volatile to most
refractory. Our best leverage on the metallicity of the DLA comes from
S, which is only mildly depleted in a dusty medium
\citep[e.g.,][]{DeCia2016} and suggests Z $\approx 0.1$Z$_{\odot}$. In
contrast, we find [Fe/H] $= -1.60 \pm 0.07$ and [Ca/H] $= -3.00 \pm
0.05$, suggesting dust depletion of the most refractory elements. The
decrement in [Fe/H] compared to [S/H] is consistent with the Galactic
halo depletion pattern of \citet{Savage1996}. Due to the low second
ionization energy of Ca (11.9 eV), it is possible that we
underestimate [Ca/H] by assuming $N$(Ca) = $N$(\ion{Ca}{2}). However,
the presence of dust depletion is clear from the under-abundance of Fe
alone. Applying the method of \citet{2021Natur.597..206D}, we use the
well-constrained S and Fe abundances to determine a total,
dust-corrected metallicity of [X/H] $= -0.8 \pm 0.1$ using the
depletion patterns of \citet{DeCia2016}. This is consistent with being
drawn from the metallicity distribution of known DLAs at low redshift
\citep{2019ApJ...887....5L}.

\subsection{An isolated dwarf galaxy pair as DLA host}
\label{sec:dwarfs}

\begin{deluxetable*}{ccc}
  \tablecaption{Properties of the dwarf galaxy hosts of the $z = 0.026$ DLA}
  \tablehead{
    \colhead{} &
    \colhead{G1} &
    \colhead{G2}
}
  \startdata
  Name & CUBS2339z002\_G6 & CUBS2339z002\_G32 \\
  RA & 23h39m14.11s & 23h39m18.82s \\
  Decl. & $-55$d23m41.9s & $-55$d24m30.5s \\
  $\Delta\,\alpha$ ($''$)\tablenotemark{a} & $+7.6$ & $+47.7$ \\
  $\Delta\,\delta$ ($''$)\tablenotemark{a} & $+9.0$ & $-39.7$ \\
  $\theta$ ($''$)\tablenotemark{a} & $11.8$ & $62.0$ \\
  $d$ (kpc)\tablenotemark{a} & $6.2$ & $32.5$ \\
  $z$\tablenotemark{b} & $0.02608$ & $0.02604$ \\
  $\Delta\,\varv_{g}$ (km\,s$^{-1}$)\tablenotemark{c} & $11.2 \pm 0.1$ & $0.4 \pm 0.1$ \\
  $m_{r}$ (mag)\tablenotemark{d} & $18.27 \pm 0.3$ & $18.31 \pm 0.3$ \\
  $M_{r}$ (mag) & $-17.01 \pm 0.3$ & $-16.97 \pm 0.3$ \\
  log$(M_{\text{star}}/\text{M}_{\odot})$\tablenotemark{e} & $8.5 \pm 0.2$ & $8.5\pm0.2$ \\
  SFR(H$\alpha$) $(M_{\odot}$ yr$^{-1})$ & $\approx 0.04$ & $\approx 0.03$ \\
  $M$(\hi) $(M_{\odot})$ & $\approx 7 \times 10^{8}$ & $\approx 7 \times 10^{8}$\\
  $M$(\mbox{H\,{\sc ii}}) $(M_{\odot})$ & $\sim 10^{7}\text{--}10^{8}$ & $\sim 10^{7}\text{--}10^{8}$ \\
  \enddata
  \tablenotetext{a}{RA offsets ($\Delta\,\alpha$), decl. offsets ($\Delta\,\delta$), angular separations ($\theta$), and impact parameters ($d$) are measured with respect to the QSO position at RA, decl. $=$ 23h39m13.22s, $-55$d23m50.8s.}\tablenotetext{b}{Light-weighted mean redshift.}\tablenotetext{c}{Velocity offset with respect to $z_{\text{DLA}} = 0.02604$.}\tablenotetext{d}{$r$-band apparent magnitudes are from DES \citep{Abbott2018} and are corrected for internal extinction adopting a characteristic $E(B-V) = 0.05\ \text{and}\ 0.1$ for G1 and G2, respectively (see \S\ref{sec:ionized_gas_metallicity}).}\tablenotetext{e}{Determined using the $M_{r} - M_{\text{star}}$ relation of \citet{Liang2014} and stellar population modeling from \texttt{bagpipes} \citep{Carnall2018}.}
  \label{tab:dwarfs}
\end{deluxetable*}

\begin{figure*}
\centering
\includegraphics[scale = 1.3]{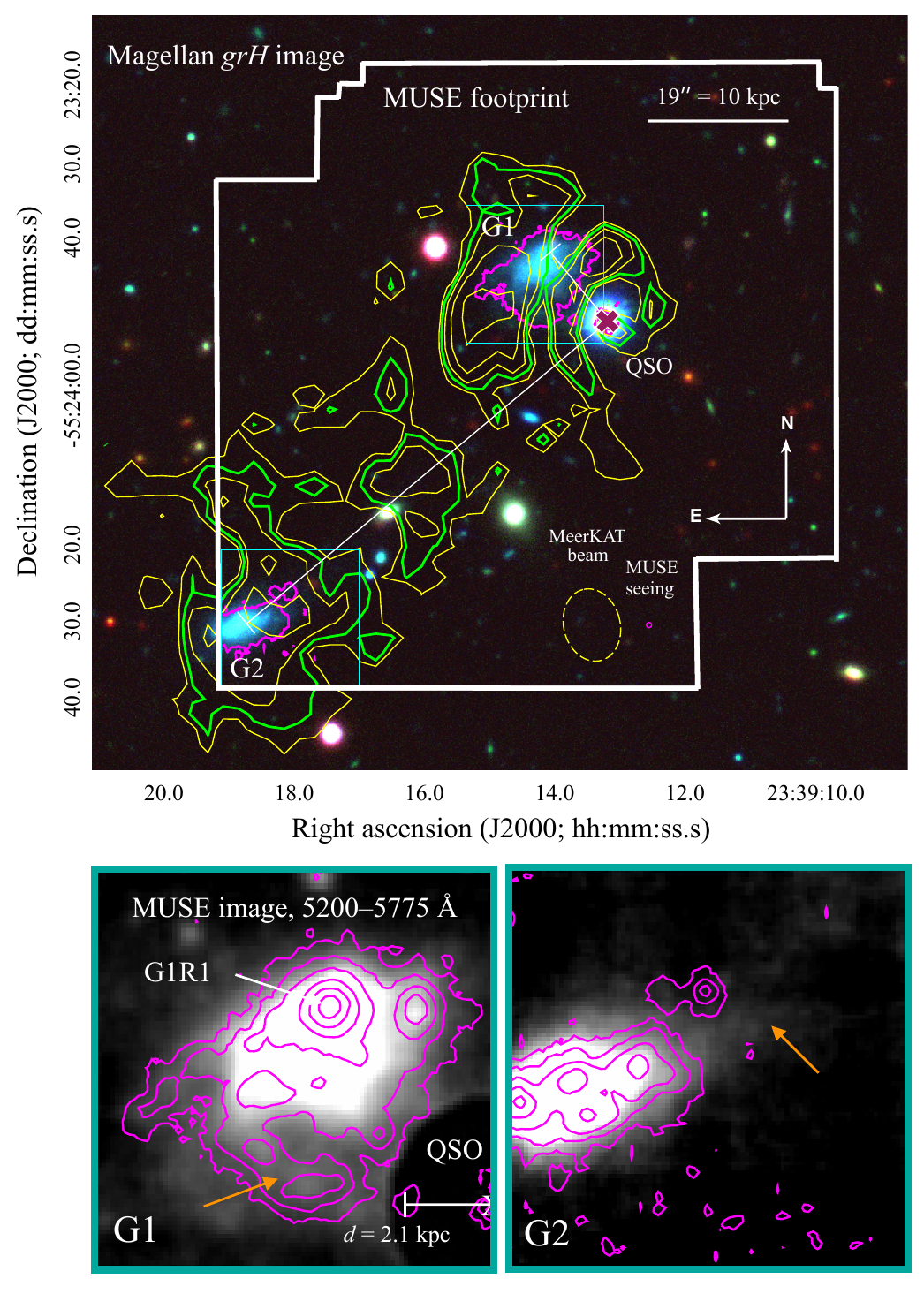}
\caption{Multi-wavelength view of the DLA host. At \textbf{top}, a
  Magellan \textit{grH} image shows two blue dwarf galaxies at the
  redshift of the DLA at impact parameters of $d = 6.2$ kpc (G1) and
  $d = 32.5$ kpc (G2). We overlay \hi\ column density contours of
  $5\times10^{19}\times$(1, 8, 16, 32)\,\cmsq\ from the MeerKAT
  observations in yellow, and we indicate the DLA column density
  threshold, $2\times10^{20}$\,\cmsq, in green. The average $3\sigma$
  column density sensitivity across the map is $3.2 \times 10^{20}$
  cm$^{-2}$ \citep[e.g.,][]{2018ApJ...865...26D}. The MeerKAT beam is
  shown in yellow at the lower right. We also overlay the 3$\sigma$
  H$\alpha$ contours from MUSE in magenta, which correspond to a
  surface brightness of $2.6 \times
  10^{-18}$~erg~s$^{-1}$~cm$^{-2}$~arcsec$^{-2}$ and $3.8 \times
  10^{-18}$~erg~s$^{-1}$~cm$^{-2}$~arcsec$^{-2}$ for G1 and G2,
  respectively. At \textbf{bottom}, we show additional H$\alpha$
  contours at the 3, 10, 30, 100, 300, and 1000$\sigma$ level on a
  white-light image constructed from the MUSE cube between
  $5200\text{--}5775$ \AA\ (chosen to avoid emission lines). The
  H$\alpha$ emission is not detected at the QSO location; the contour
  shown at the QSO position is a result of residuals from the QSO PSF
  subtraction. However, an H$\alpha$ blob is detected at an impact
  parameter of $d = 2.1$ kpc. We label the brightest star-forming
  region in G1 as G1R1. The stellar continuum shows evidence of
  disturbance at its faint outskirts, including possible stellar
  streams pointing towards the QSO as indicated by orange arrows. The
  H$\alpha$ features at the bottom of the G2 frame are due to elevated
  noise at the edge of the MUSE detector.}
\label{fig:pointing}
\end{figure*}

We observe two dwarf galaxies at the redshift of the DLA and employ a
deep galaxy redshift survey to characterize their
environment. \citet{Chen2020} describe the redshift survey conducted
in all CUBS fields with LDSS-3C and IMACS on Magellan and
VLT-MUSE. The spectroscopic component of this survey targets galaxies
fainter than 18th magnitude within $\theta \lesssim 10'$, so we
primarily use photometric redshifts from the Dark Energy Survey (DES;
\citealt{Abbott2018}). At the redshift of the DLA, a blue galaxy with
$L = 0.01L_{*}$ has an apparent $r$-band magnitude of $m_{r} = 19.2$
assuming $M_{r,*} = -21.1$ \citep{Cool2012}, where $M_{r,*}$ is the
characteristic rest-frame absolute $r$-band magnitude at the break of
the luminosity function. We estimate the uncertainty on these
photometric redshifts using 16 galaxies with $m_{r} < 19$ with robust
spectroscopic redshifts within $\theta = 11'$. For galaxies at $z <
0.3$, we find $dz/(1 + z) < 0.02$. We do not find any galaxies with
$|z_{\text{phot}} - z_{\text{DLA}}| < 0.02$. There are two galaxies
with $|z_{\text{phot}} - z_{\text{DLA}}| = 0.02 - 0.04$. The first is
at $z_{\text{phot}} = 0.055$ and $d = 95$ kpc; this system has $m_{r}
= 18.4$. The second is at $z_{\text{phot}} = 0.064$ and $d = 255$ kpc
and has $m_{r} = 17.4$. Thus, aside from the two dwarf galaxies, there
are no galaxies more luminous than $\approx 0.05L_{*}$ within $d
\approx 350$ kpc.

As shown in Fig.~\ref{fig:pointing} and Table~\ref{tab:dwarfs}, the
two dwarf galaxies are separated by 33 kpc and are within a projected
velocity of $\Delta\,\varv_{g} \lesssim 10$ km~s$^{-1}$ of the DLA. We
refer to these galaxies as G1 and G2, where G1 is the galaxy at
smaller impact parameter, $d$, with respect to the QSO. At $d = 6.2$
kpc, G1 is projected in close enough proximity to the QSO that the
$3\sigma$ optical continuum light overlaps with the wing of the QSO
point spread function (PSF) on the southwest side of the galaxy. G2 is
found at the edge of the MUSE footprint at $d = 32.5$ kpc and $\approx
15$\% of the area within the $3\sigma$ $r$-band contour is found
outside of the IFU coverage.

We estimate the stellar mass of both galaxies to be
log$(M_{\text{star}}/\text{M}_{\odot}) = 8.5 \pm 0.2$ from
extinction-corrected, $r$-band photometry from DES using the $M_{r} -
M_{\text{star}}$ relation derived from NASA–Sloan Atlas galaxies by
\citet{Liang2014}. We find comparable mass estimates from stellar
population synthesis models using \texttt{bagpipes}. We estimate the
halo mass of the galaxies to be log$(M_{\text{h}}/\text{M}_{\odot}) =
10.8 \pm 0.2$ from the stellar mass-halo mass relation for field
dwarfs determined by \citet{Read2017}; this is consistent with the
models of \citet{Behroozi2019}. Following \citet{Maller2004}, this
yields a virial radius of $R_{\text{vir}} \approx 100_{-10}^{+20}$
kpc. Here, $R_{\text{vir}}$ is defined based on the overdensity
condition of \citet{Bryan1998} at the redshift of the dwarfs. The
projected separation of the dwarfs is thus approximately one third of
their virial radii. As shown in Fig.~\ref{fig:pointing}, the optical
continuum morphology of the faint outskirts of both dwarfs show
disturbances; these irregularities are most prominent on the sides of
the galaxies closest to their companion, including a faint stellar
stream protruding from G1 that points toward the QSO. The
extinction-corrected H$\alpha$ luminosity of G1, $L(\text{H}\alpha)
\approx 8 \times 10^{39}$ erg s$^{-1}$, suggests a star-formation rate
(SFR) of $\approx 0.04$ M$_{\odot}$ yr$^{-1}$ following the relation
of \citet{Calzetti2008}. About two thirds of the star formation is
found in the bright, star-forming knot labeled G1R1 in
Fig.~\ref{fig:pointing}. G2 is $\approx 40\%$ dimmer than G1 in
H$\alpha$ and thus has SFR(H$\alpha$) $\approx 0.03$ M$_{\odot}$
yr$^{-1}$ (a small fraction of the H$\alpha$ flux may be unaccounted
for due to the proximity of the galaxy to the edge of the detector).

\subsection{Neutral gas properties of DLA host}
\label{sec:HI}

We detect spatially extended \hi\ 21-cm emission that reveals a
neutral gas envelope surrounding G1 and G2 on a scale of $> 40$ kpc,
including an apparent tidal bridge between them. We display the
\hi\ column density contours in Fig.~\ref{fig:pointing}. The bridge
has a width of $\sim20^{\prime\prime}$, or $10$ kpc, at the level of
$5\times10^{19}$\cmsq\ and is thus resolved by MeerKAT's synthesized
beam (FWHM = $4$ kpc $\times\ 3$ kpc). The total integrated line flux
of the system is $0.44 \pm 0.06$\,Jy\,\kms, which corresponds to a
total \hi\ gas mass of $M($\hi$) \approx 1.3\times
10^{9}$\,M$_{\odot}$.  If we bisect the gas distribution at the
midpoint between the galaxies, the two dwarfs are associated with
nearly the same \hi\ mass.

As shown in Fig.~\ref{fig:pointing}, the QSO sightline intersects the
outer gaseous envelope of G1. In the vicinity of the QSO,
\hi\ emission from several voxels in the image cube contributes to the
\hi\ 21-cm column density. The emission is detected in individual
spectral channels at the $2\text{--}3\sigma$ level.  Within the
MeerKAT synthesized beam, the characteristic column density at the
location of the peak flux density of the QSO is
log[$N$(\hi)/cm$^{-2}$] $\approx 20.6$.  Despite the possible impact
of beam dilution in a clumpy medium, this is remarkably consistent
with the column density determined from the \textit{HST}-COS
absorption-line spectroscopy.

J2339-5523 is unresolved (size $<$ 300\,pc) in the MeerKAT image.  We
visually examined the spectral line image cube and confirmed that no
negative features representing \hi\ 21-cm absorption at $> $2$\sigma$
level ($\int\tau d\varv$ = 0.028\,\kms) are present towards the QSO.
The strength of \hi\ 21-cm absorption depends on the \hi\ column
density as well as the spin temperature, $T_{s}$, and the covering
factor of the absorbing gas, $f_c$.  If the radio-emitting region of
the QSO is more spatially extended than the absorbing gas, then $f_c <
1$.  The \hi\ column density towards the UV-emitting region of the QSO
is well constrained through the \textit{HST}-COS observations of the
DLA. If the UV- and radio-emitting regions of the QSO are spatially
aligned, we can use the measured $N$(\hi) from COS to constrain the
ratio of the spin temperature to the covering factor. Using a
5$\sigma$ upper limit on \hi\ 21-cm absorption in the unsmoothed
MeerKAT spectrum with a spectral rms of 0.44\,mJy\,beam$^{-1}$ per
channel, we estimate a lower limit of $T_{s}/f_c > 1880$ K assuming a
FWHM of 17\,\kms\ ($b$ = 10\,\kms) for the \hi\ absorption line based
on the metal lines detected in the COS and MIKE spectra.

In the top panels of Fig.~\ref{fig:kinematics}, we show the
line-of-sight velocity and velocity dispersion of the neutral gas
traced by \hi\ 21-cm emission. The velocity shear across the neutral
gas distribution is $\approx \pm 20$ km s$^{-1}$. The characteristic
velocity dispersion within the bodies of G1 and G2 is $10$--$15$ km
s$^{-1}$, and this value is a factor of two higher in parts of the
extended medium, suggesting possible dynamical disturbance. Due to the
relatively small number of MeerKAT beams sampling the gas
distribution, we cannot draw strong conclusions about whether the
bridge gas arises from G1 or G2. It is also not clear if the gas
distribution in either galaxy is disk-like; there is no definitive
evidence for organized rotation, and it is common for the gas content
in galaxies of this mass to have significant dispersion support
\citep[e.g.,][]{2018MNRAS.477.1536E}.

\subsection{Warm ionized gas properties of DLA host}
\label{sec:ionized_gas_properties}

We analyze the warm ionized medium (WIM) in both dwarf galaxies with
the goal of comparing observable properties of the ISM with those of
the damped absorber to better characterize the origin of the neutral
gas envelope. Among the emission lines covered by MUSE are H$\alpha$,
H$\beta$, [OIII]$\lambda\lambda$4960, 5008, [NII]$\lambda\lambda$6549,
6585, [SII]$\lambda\lambda$6718, 6732, and [SIII]$\lambda$6313, 9071,
all of which are detected in at least one star-forming region in one
or both galaxies.

\subsubsection{WIM mass and morphology}
\label{sec:ionized_gas_mass}

As shown in Fig.~\ref{fig:pointing}, there is no evidence of spatially
extended, optical line emission on scales larger than the stellar
components of G1 and G2. The $3\sigma$ H$\alpha$ surface brightness
detection threshold in the vicinity of G1 (G2) is $2.6 \times
10^{-18}$ erg s$^{-1}$ cm$^{-2}$ arcsec$^{-2}$ ($3.8 \times 10^{-18}$
erg s$^{-1}$ cm$^{-2}$ arcsec$^{-2}$). Notably, the WIM does not
overlap with the quasar sightline at the detection threshold of the
data. The line-emitting feature in closest projected proximity to the
quasar sightline ($d \approx 2$ kpc) is an isolated cloud separated
from the main body of G1 in the $3\sigma$ H$\alpha$ surface-brightness
contour (see Fig.~\ref{fig:pointing}). We used an optimal extraction
method implemented in CubExtractor to search for H$\alpha$ and [NII]
emission on top of the QSO, but the noise properties of the region
affected by the bright QSO PSF prohibit a sensitive constraint on the
H$\alpha$ surface brightness at this location.

We estimate the warm ionized gas mass in the ISM of G1 and G2 as follows. For a
clumping factor $\mathcal{C} \equiv \langle n_{e}^{2} \rangle/\langle n_{e}
\rangle^{2}$, the H$\alpha$ surface brightness depends on the mean
electron density, $\langle n_{e} \rangle$, and the pathlength through
the gas, $L$, according to:
\begin{equation}
  I(\text{H}\alpha) \approx 2 \times 10^{-15}\ \frac{\mathcal{C} \langle n_{e} \rangle^{2}}{(1 + z)^{4}}\ \frac{L}{\text{kpc}}\ \text{erg}\ \text{s}^{-1}\ \text{cm}^{-2}\ \text{arcsec}^{-2}.
  \label{eq:em_measure}
\end{equation}
Here the electron density is in units of cm$^{-3}$, and we assume an
electron temperature $T_{e} = 10^{4}$ K. If the ionized gas is found
in a relatively thin disk ($L \approx 1$ kpc) in G1, then
$M$(\mbox{H\,{\sc ii}}) $\approx 4\text{--}10 \times 10^{7}$
M$_{\odot}$ for $\mathcal{C}$ between 10 and 1. If instead $L$ is
closer to the projected size of the line-emitting region ($L \approx
7$ kpc), then the ionized gas mass may be as high as $0.9\text{--}3
\times 10^{8}$ M$_{\odot}$ for the same range of clumping factor. At
approximately half of the H$\alpha$ luminosity, the ionized gas mass
estimates for G2 are roughly half of those for G1. Across these
models, the electron density varies between $\langle n_{e} \rangle
\approx 0.005\text{--}1$ cm$^{-3}$, which is consistent with the
constraint from $I$([SII]$\lambda$6718)/$I$([SII]$\lambda$6732) that
$n_{e} \lesssim 10^{2}$ cm$^{-3}$
\citep[e.g.,][]{Osterbrock2006}. Thus, the warm ionized phase of the
ISM likely contributes between a few percent to 50\% as much mass as
the neutral phase.

\subsubsection{WIM kinematics}
\label{sec:ionized_gas_kinematics}

\begin{figure*}
\centering
\includegraphics[scale = 1.0]{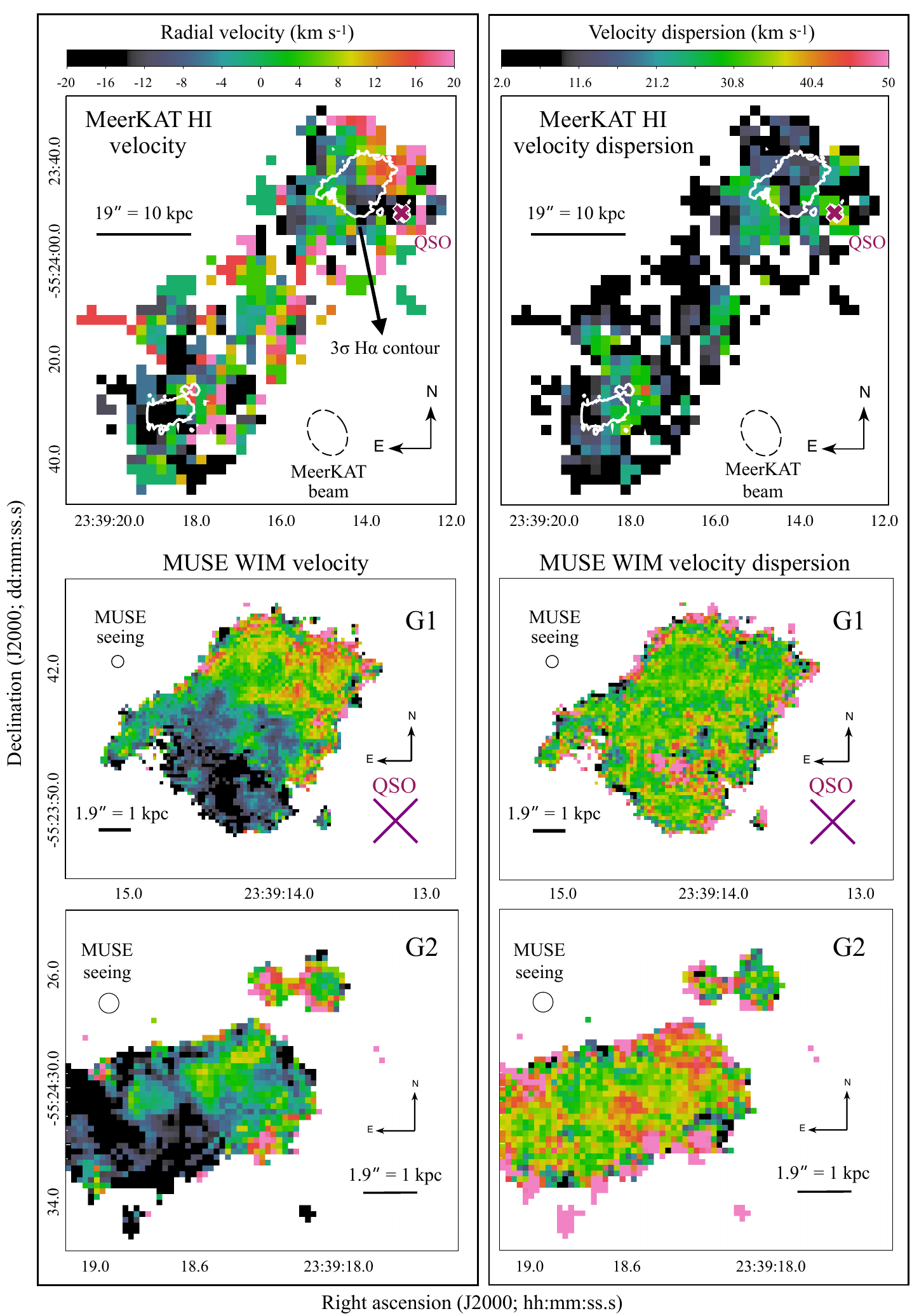}
\caption{\textbf{At left,} the radial velocities of the neutral gas
  traced by \hi\ 21-cm emission \textbf{(top)} and ionized gas
  observed via optical line emission \textbf{(bottom)} show notable
  consistency with the velocity of the DLA, where the velocity
  zeropoint is set by the DLA redshift ($z = 0.026$). \textbf{At
    right}, we show the velocity dispersion of the multi-phase gas. In
  the bodies of the dwarf galaxies, the characteristic dispersion is
  $10$--$15$ km s$^{-1}$ in the neutral gas and $30$ km s$^{-1}$ in
  the ionized gas. While this hints at modest turbulence in the latter
  phase, the optical emission lines are largely unresolved (the MUSE
  spectral resolution at H$\alpha$ has a FHWM of $\approx 113$ km
  s$^{-1}$, or $\sigma \approx 48$ km s$^{-1}$). The isolated clumps
  on the southern side of G2 are marginal detections with uncertain
  line widths.}
\label{fig:kinematics}
\end{figure*}

We show the generally quiescent kinematics of the warm ionized gas in
the dwarf galaxies in Fig.~\ref{fig:kinematics}. In G1, the
line-of-sight velocities generally fall in the range $-25 $ km s$^{-1}
\le \varv_{\text{rad}} \le +25$ km s$^{-1}$; an ordered pattern is
evident, with the southern side of the galaxy approaching and the
northern side receding. The discrete cloud closest in projection to
the QSO has a median $\varv_{\text{rad}} = -4 \pm 10$ km s$^{-1}$, and
comparable velocity offsets with respect to the absorber are seen in
the portion of the main body of the galaxy in closest projected
proximity. We find similar results for G2, with milder evidence for an
ordered velocity gradient of a few tens of km s$^{-1}$ from the
southeast (approaching) to northwest (receding) side. The
characteristic velocity dispersion of the WIM in both galaxies is
$\sigma \equiv \text{FWHM}/(2\sqrt{2\text{ln}(2)}) = 30\text{--}40$ km
s$^{-1}$ (corrected for spectral resolution), which is not resolved at
the spectral resolution of the MUSE instrument. There is marked
consistency between the velocity centroids of the DLA, and the
\hi\ and \mbox{H\,{\sc ii}} gas.

\subsubsection{WIM metallicity and dust content}
\label{sec:ionized_gas_metallicity}

\begin{figure*}
\includegraphics[scale = 0.8]{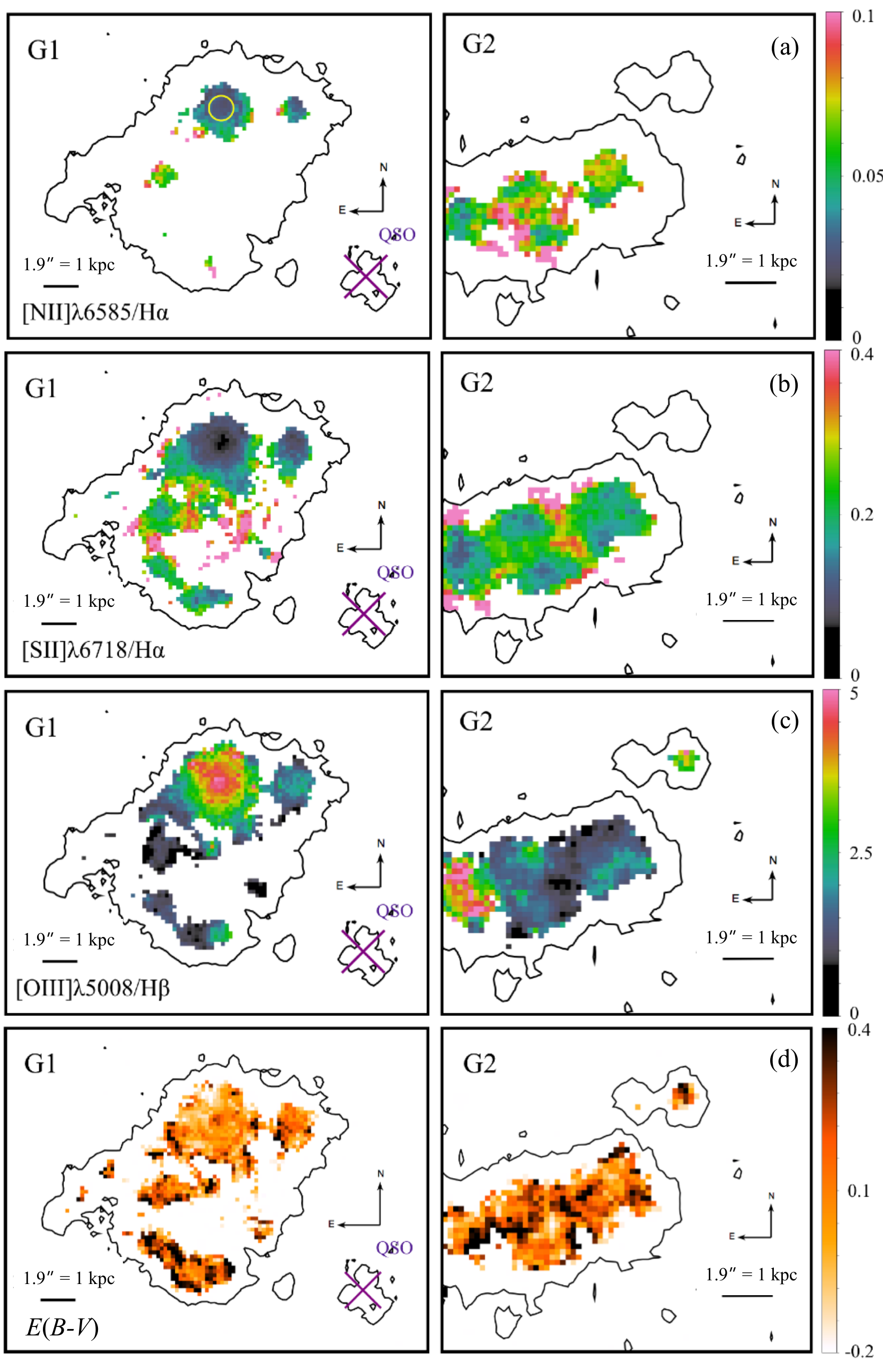}
\centering
\caption{Maps of G1 (left) and G2 (right), showing the emission-line
  intensity ratios \textbf{(a)}
  $I$([NII]$\lambda$6585)/$I$(H$\alpha$), \textbf{(b)}
  $I$([SII]$\lambda$6718)/$I$(H$\alpha$), and \textbf{(c)}
  $I$([OIII]$\lambda$5008)/$I$(H$\beta$). In panel \textbf{(d)}, we
  show the reddening, $E(B-V)$, as determined from the Balmer
  decrement, indicating the presence of dust in the ISM. The included
  spaxels have detections of both relevant lines at the $\ge 3\sigma$
  level. The black contour is the 3$\sigma$ H$\alpha$ detection
  threshold. The contour shown at the QSO location is a result of
  residuals from the QSO PSF subtraction. In the left panels, the
  stellar stream pointing toward the QSO from G1 is seen to be
  accompanied by warm ionized gas that contains dust. The aperture
  indicated in yellow at the top left is the region where we detect
  both the auroral [SIII]$\lambda$6313 and the nebular
  [SIII]$\lambda$9071 lines that permit a direct measurement of
  $T_{e}$ and thus [S/H].}
\label{fig:line_ratios}
\end{figure*}

For G1R1, we detect both the auroral [SIII]$\lambda$6313 and the
nebular [SIII]$\lambda$9071 transitions, permitting a measurement of
$T_{e}$ and thus our most direct constraint on the gas-phase
metallicity. In the upper left panel of Fig.~\ref{fig:line_ratios}, we
show the aperture within which both lines are detected at the
3$\sigma$ level, and we find $I$([SIII]($\lambda$9533 +
$\lambda$9071))/$I$(SIII]$\lambda$6313) $= 29.0 \pm 0.9$ within this
  region (note that the nebular lines have a fixed ratio of
  $I$([SIII]$\lambda$9533)/$I$([SIII]$\lambda$9071) $=
  2.44$). Following \citet{Osterbrock2006}, this yields $T_{e} \approx
  1.4 \times 10^{4}$ K at the density constrained by the [SII] doublet
  ($n_{e} \lesssim 10^{2}$ cm$^{-3}$). Adopting this value of $T_{e}$,
  we use the $I$([\ion{S}{2}]$\lambda$6718)/$I$(H$\alpha$) and
  $I$([\ion{S}{3}]$\lambda$9533)/$I$(H$\alpha$) line ratios to
  estimate S/H = S$^{+}$/H + S$^{++}$/H $\approx 10\text{--}20$\% of
  the solar value, consistent with the abundance determined for the
  DLA in \S\ref{sec:DLA_prop}.

We additionally detect several strong-line metallicity indicators in
multiple star-forming regions in both galaxies, including N$_{2} =$
$I$([\ion{N}{2}]$\lambda$6585)/$I$(H$\alpha$)
\defcitealias{2004MNRAS.348L..59P}{PP04}\citep[e.g.,][hereafter
  \citetalias{2004MNRAS.348L..59P}]{2004MNRAS.348L..59P} (see
Fig.~\ref{fig:line_ratios}). Following
\citetalias{2004MNRAS.348L..59P}, we find 12 + log(O/H) $\approx 8.0$
(20\% solar) from N$_{2}$ in G1R1 (note the 0.2 dex systematic
uncertainty in the \citetalias{2004MNRAS.348L..59P} relation). We
observe values up to 12 + log(O/H) = 8.3--8.4 (40--50\% solar) in
lower surface brightness regions of this galaxy, but we regard this as
an upper limit due to the likely presence of diffuse ionized gas that
elevates the observed line ratios and thus biases the metallicity
measurement \citep[e.g.,][]{2017MNRAS.466.3217Z,
  2019MNRAS.489.4721V}. In the star-forming regions of G2, 12 +
log(O/H) is generally $\approx 0.1$ dex higher than in G1, suggesting
slightly elevated metallicity in this galaxy. We corroborate a
metallicity of a few tens of percent solar for G1 and G2 using
\texttt{Cloudy} photoionization models to reproduce the observable
emission-line ratios \citep{Ferland2017}. The dust extinction maps
shown in the bottom panel of Fig.~\ref{fig:line_ratios} indicate a
characteristic $E(B-V) = 0.05\ \text{and}\ 0.1$ for G1 and G2,
respectively. This is comparable to the characteristic $E(B-V) \approx
0.1$ found for galaxies of this mass by \citet{2018ApJ...859...11S},
suggesting that G1 and G2 are typical in their dust content.

\section{Discussion and Concluding Remarks}
\label{sec:diss}

This case study clearly confirms an association between the $z =
0.026$ DLA and a neutral gas envelope surrounding two interacting
dwarf galaxies in a rare instance of spatially resolved \hi\ 21-cm
emission that is coincident with a UV-selected DLA outside of the
optical extent of a galaxy\footnote{See \citet{2002A&A...388..383C}
  for the one known instance inside of the optical extent.}. Both the
ISM of G1 and the DLA have relatively low metallicities ($Z \approx
0.1Z_{\odot}$), contain dust, and show comparable kinematics,
suggesting that the gas in the envelope originated in the ISM of the
dwarf galaxies. Opportunities to compare the chemical enrichment of
DLAs with that of the ionized ISM of the host galaxies remain
rare. \citet{2003ApJ...595..760A} use a down-the-barrel approach to
demonstrate that the neutral ISM has $\alpha$-element abundances below
that of the \hii\ regions in I Zw 18, implying that the star-forming
regions are enriched by recent star formation. In contrast, the DLA
galaxy SBS 1543+593 has consistent $\alpha$-element abundances in its
\hii\ regions and in the damped absorber that coincides with its inner
ISM \citep{2004ApJ...600..613S, 2005ApJ...625L..79S,
  2005ApJ...635..880B}. The J2339-5523 DLA provides a second case of
consistency between the $\alpha$-element abundances of the damped
absorber and the \hii\ regions in the host galaxy. Establishing a
statistical sample of such systems in the local Universe will shed
light on the nature of the host galaxies of low-metallicity DLAs at
higher redshift.

Correlations between galaxy mass and/or morphology, optical size, and
\hi\ size are well characterized and establish expectations for the
extent of the neutral gas distributions around the dwarf
galaxies. \citet{1994AJ....107.1003C} report mean \hi\ to optical
diameter ratios in the range $1.5$--$1.9$ for disk galaxies with a
variety of morphological types in the field, while
\citet{1997A&A...324..877B} find a consistent mean value of $1.7 \pm
0.5$ independent of morphological type. The equivalent ratio for G1
and G2 is between 2 and 3 and thus falls $1$--$2\sigma$ towards the
upper end of the distribution. Additionally, the LITTLE THINGS
\hi\ 21-cm survey of nearby dwarf irregular galaxies
\citep{2012AJ....144..134H}, as well as dwarf galaxy samples from
\citet{2009MNRAS.400..743K} and \citet{2014A&A...566A..71L}, suggest a
relatively flat distribution of \hi\ to optical diameter ratio, with
some systems in the range of $2$--$4$ \citep{2017ASSL..434..209B}.

\citet{2016MNRAS.460.2143W} used a large sample of more than 500
galaxies from various samples to demonstrate a tight relation between
\hi\ diameter, $D$(\hi), and $M$(\hi). The relation suggests an
overall similarity in the evolution of gas-rich galaxies, from small
dwarfs to large spirals.  The observed neutral gas mass is $M$(\hi)
$\approx 7 \times 10^{8}$ M$_{\odot}$ for both dwarfs if we bisect the
total \hi\ distribution at their midpoint. This corresponds to a size
of $D$(\hi) $\approx 16$ kpc based on the \citet{2016MNRAS.460.2143W}
relation.  For G1, the extent of the \hi\ is 14 kpc at a gas surface
density of $\Sigma$(\hi) = 1 $M_\odot$\,pc$^{-2}$, well within the
$3\sigma$ scatter of the relation. In summary, excluding the bridge,
the spatial extent of the observed \hi\ distribution around G1 and G2
is consistent with the known \hi\ size-mass relation. This is in line
with the demonstrated robustness of this relation in the presence of
environmental processes that cause a truncation of the
\hi\ distribution but leave the \hi\ size-mass relation unaffected
\citep[][]{Stevens19}.

Following \citet{King1962}, we characterize the impact of tidal
interaction on the dwarf galaxies. For a
body, $M_{1}$, on a circular orbit in the gravitational field of
$M_{2}$, the radius beyond which tidal stripping will occur, $r_{t}$,
can be expressed as:
\begin{equation}
  r_{t} = r_{\text{sep}} \left( \frac{M_{1}}{3M_{\text{2,encl}}} \right)^{1/3}.
\end{equation}
Here, $r_{\text{sep}}$ is the distance between the objects at closest
approach and $M_{\text{2,encl}}$ is the mass of $M_{2}$ enclosed
within $r_{\text{sep}}$. We assume that the current projected
separation of the dwarf galaxies is their distance of closest approach
($r_{\text{sep}} = 33$ kpc) and calculate $M_{\text{2,encl}}$
including contributions from stars, gas, and a dark matter halo
described by a Navarro-Frenk-White profile \citep{Navarro1997} with a
concentration parameter of $c = 10 - 20$
\citep[e.g.,][]{Zhao2009}. Taking $M_{1}$ to be the stellar, gas, and
dark matter mass contained within the stellar body, we find that the
truncation radius is in the range $r_{t} \approx 11 - 13$ kpc. This
exceeds the observed value of $\approx 7$ kpc, which suggests either
that the galaxies are intrinsically small enough to avoid tidal
truncation, or that a closer approach between the dwarf galaxies in
the past is responsible for truncating them to their current size. The
latter scenario is supported by the presence of the \hi\ bridge and
the faint irregularities observed in the outskirts of the stellar and
ionized gas distributions of both galaxies. The observed offset in the
centers of the stellar and neutral gas distributions in G1 may also
indicate tidal disruption of the gaseous component of the dwarf. All
evidence considered, it is likely that the DLA arises from gas that
originated in the ISM of G1 and has since been perturbed by tidal
interaction.

The definitive lack of a massive companion allows us to isolate the
effects of dwarf-dwarf interactions on their gaseous reservoirs in the
absence of processes such as ram-pressure stripping in a massive, hot
halo. In this system, the cross-section for producing damped absorbers
is much larger than the inner ISM alone. Gas with $N$(\hi) $\ge 2
\times 10^{20}$ cm$^{-2}$ is seen in the ISM, the outskirts of the
\hi\ distributions, and the adjoining bridge, informing the diversity
of environments that can host DLAs. It is likely that the
cross-section for ionized absorbers is larger still, but below the
MUSE detection threshold; the $z = 0.364$ Lyman limit system (LLS)
detected in CUBS towards J0248$-$4048 arises from an analogous system
of a relatively isolated dwarf galaxy pair at an impact parameter of
several tens of kpc. The metallicity and chemical abundance pattern of
the LLS suggests that it arises in tidal feature(s) composed of former
ISM material \citep{2021MNRAS.506..877Z}. The Magellanic Stream
appears to be a local example of a multiphase feature whose formation
was dominated by tidal forces from the Large Magellanic Cloud acting
on its smaller companion on initial infall into the halo of the Milky
Way \citep[e.g.,][]{2012MNRAS.421.2109B}. Thus, models of the
evolution of dwarf galaxy pairs --- both within and outside of the
halos of massive galaxies --- must account for the role of dwarf-dwarf
interactions in producing tidal features that enhance the
cross-section for neutral absorbers.

Our comprehensive characterization of the galactic origin of the $z =
0.026$ DLA was only made possible by the joint analysis of
\textit{HST} UV absorption-line and MeerKAT \hi\ 21-cm emission-line
data alongside detailed galaxy spectroscopy. As we enter further into
the era of the Square Kilometer Array (SKA) and the next generation
Very Large Array (ngVLA) with much higher sensitivity and spatial
resolution, leveraging the time-limited availability of space-based UV
resources in conjunction with spatially-resolved 21-cm mapping will be
critical to enhance our understanding of the diverse galactic
environments that host the majority of the neutral gas in the
Universe.

\acknowledgments We thank Tom Cooper for useful discussions and for
his valuable contributions to the CUBS galaxy redshift survey, Zhijie
Qu for helpful comments on the paper draft, and Alex Drlica-Wagner for
assistance in extracting photometric data from the Dark Energy Survey
catalog. We thank the anonymous referee for constructive comments that
improved the presentation of the paper.

EB and HWC acknowledge partial support from HST-GO-15163.001A and NSF
AST-1715692 grants. SC gratefully acknowledges support from the
European Research Council (ERC) under the European Union’s Horizon
2020 research and innovation programme grant agreement No 864361. KLC
acknowledges partial support from NSF AST-1615296. CAFG was supported
by NSF through grants AST-1715216, AST-2108230, and CAREER award
AST-1652522; by NASA through grant 17-ATP17-0067; by STScI through
grant HST-AR-16124.001-A; and by the Research Corporation for Science
Advancement through a Cottrell Scholar Award. JKK acknowledges support
by the Swiss National Science Foundation under grant 185692. SL was
funded by FONDECYT grant number 1191232. FSZ is grateful for the
support of a Carnegie Fellowship from the Observatories of the
Carnegie Institution for Science. This material is based upon work
supported by NASA under award number 80GSFC21M0002.

This work is based on observations made with ESO Telescopes at the
Paranal Observatory under programme ID 0104.A-0147(A), observations
made with the 6.5m Magellan Telescopes located at Las Campanas
Observatory, and spectroscopic data gathered under the
HST-GO-15163.01A program using the NASA/ESA Hubble Space Telescope
operated by the Space Telescope Science Institute and the Association
of Universities for Research in Astronomy, Inc., under NASA contract
NAS 5-26555. The MeerKAT telescope is operated by the South African
Radio Astronomy Observatory, which is a facility of the National
Research Foundation, an agency of the Department of Science and
Innovation. The MeerKAT data were processed using the MALS computing
facility at IUCAA (\url{https://mals.iucaa.in/releases}).

This project used public archival data from the Dark Energy Survey
(DES). Funding for the DES Projects has been provided by the
U.S. Department of Energy, the U.S. National Science Foundation, the
Ministry of Science and Education of Spain, the Science and Technology
Facilities Council of the United Kingdom, the Higher Education Funding
Council for England, the National Center for Supercomputing
Applications at the University of Illinois at Urbana-Champaign, the
Kavli Institute of Cosmological Physics at the University of Chicago,
the Center for Cosmology and Astro-Particle Physics at the Ohio State
University, the Mitchell Institute for Fundamental Physics and
Astronomy at Texas A\&M University, Financiadora de Estudos e
Projetos, Funda{\c c}{\~a}o Carlos Chagas Filho de Amparo {\`a}
Pesquisa do Estado do Rio de Janeiro, Conselho Nacional de
Desenvolvimento Cient{\'i}fico e Tecnol{\'o}gico and the
Minist{\'e}rio da Ci{\^e}ncia, Tecnologia e Inova{\c c}{\~a}o, the
Deutsche Forschungsgemeinschaft, and the Collaborating Institutions in
the Dark Energy Survey.

The Collaborating Institutions are Argonne National Laboratory, the
University of California at Santa Cruz, the University of Cambridge,
Centro de Investigaciones Energ{\'e}ticas, Medioambientales y
Tecnol{\'o}gicas-Madrid, the University of Chicago, University College
London, the DES-Brazil Consortium, the University of Edinburgh, the
Eidgen{\"o}ssische Technische Hochschule (ETH) Z{\"u}rich, Fermi
National Accelerator Laboratory, the University of Illinois at
Urbana-Champaign, the Institut de Ci{\`e}ncies de l'Espai (IEEC/CSIC),
the Institut de F{\'i}sica d'Altes Energies, Lawrence Berkeley
National Laboratory, the Ludwig-Maximilians Universit{\"a}t
M{\"u}nchen and the associated Excellence Cluster Universe, the
University of Michigan, the National Optical Astronomy Observatory,
the University of Nottingham, The Ohio State University, the OzDES
Membership Consortium, the University of Pennsylvania, the University
of Portsmouth, SLAC National Accelerator Laboratory, Stanford
University, the University of Sussex, and Texas A\&M University.

Based in part on observations at Cerro Tololo Inter-American
Observatory, National Optical Astronomy Observatory, which is operated
by the Association of Universities for Research in Astronomy (AURA)
under a cooperative agreement with the National Science
Foundation. The National Radio Astronomy Observatory is a facility of
the National Science Foundation operated under cooperative agreement
by Associated Universities, Inc. This research has made use of NASA’s
Astrophysics Data System and the NASA/IPAC Extragalactic Database
(NED) which is operated by the Jet Propulsion Laboratory, California
Institute of Technology, under contract with the National Aeronautics
and Space Administration. This work made use of the KUBEVIZ software
which is publicly available at
\url{http://www.mpe.mpg.de/~dwilman/kubeviz/}.

\software{\texttt{ARTIP} \citep{Gupta21}, \texttt{bagpipes} \citep{Carnall2018}, \texttt{CalCOS}, \texttt{CarPy} \citep{Kelson2003}, \texttt{CLOUDY} \citep{Ferland2013}, \texttt{CubExtractor} \citep{2019MNRAS.483.5188C}, \texttt{KUBEVIZ}, \texttt{SOFIA} \citep{Serra15, Westmeier21}}





\end{document}